\documentclass[aps,prl,twocolumn,floatfix,10pt]{revtex4-2}

\usepackage{graphicx}
\usepackage{scrextend}
\usepackage{algorithm,algpseudocode}
\usepackage[looser,scaled=1.05]{newtxtext} 
\usepackage[vvarbb,subscriptcorrection,scaled=1.05]{newtxmath} 
\usepackage{hyperref}

\hypersetup{
    colorlinks=true,
    linkcolor=blue,
    citecolor=blue,
    filecolor=blue,
    urlcolor=blue,
}

\DeclareMathAlphabet{\mathcal}{OMS}{cmsy}{m}{n}

\begin{document}
\title{\vspace*{0.51cm}Low-Depth Quantum Metropolis Algorithm}
\author{Jonathan E. Moussa}
\affiliation{Molecular Sciences Software Institute, Virginia Tech, Blacksburg, Virginia 24060, USA}
\email{godotalgorithm@gmail.com}

\begin{abstract}
\vspace*{0.35cm}
We construct a simple quantum version of the\vspace*{1.4pt} classical Metropolis algorithm to prepare and observe quantum thermal states.
It induces both a quantum\vspace*{1.4pt} Markov chain that mixes the quantum thermal state and a classical Markov chain
 that mixes its observable\vspace*{1.4pt} measurements and enables a low-depth quantum circuit implementation.
A Gaussian-filtered variant of quantum\vspace*{1.4pt} phase estimation enables thermalization times
 proportional to the reciprocal\vspace*{1.4pt} of temperature and the logarithm of biasing error.
 This matches the thermalization time of imaginary-time evolution, against which our algorithm performs favorably.\vspace*{1.4pt}
\vspace*{1.4cm}
\end{abstract}

\maketitle

We expect universal, efficient, unbiased simulations of quantum thermal states
 to require quantum algorithms on quantum computers \cite{quantum_simulation}.
Following the great successes of classical Markov chain Monte Carlo (MCMC) algorithms
 in sampling classical thermal states on classical computers \cite{MCMC_paradigm},
 a natural progression is to consider quantum MCMC algorithms \cite{quantum_thermalization} for this purpose.
The simplest approach is to extend an established classical MCMC algorithm
 such as the Metropolis algorithm \cite{classical_Metropolis} to the quantum case.

The essential feature of classical MCMC algorithms is passive, unbiased sampling of thermal expectation values.
While there have already been two proposals for quantum Metropolis algorithms \cite{quantum_Metropolis,quantum_quantum_Metropolis},
 neither retains this feature.
To prepare a quantum thermal state, $\rho_\beta = e^{-\beta H}/\mathrm{tr}(e^{-\beta H})$, at a temperature $\beta^{-1}$,
 we need some form of access to the energies $E_a$ and stationary states $|\psi_a\rangle$ of its Hamiltonian,
\begin{equation}
 H = \sum_{a} E_a |\psi_a \rangle \langle \psi_a| .
\label{hamiltonian}
\end{equation}
Both algorithms use quantum phase estimation (QPE) \cite{QPE} to probe $E_a$ and $|\psi_a\rangle$, assuming infinite energy resolution to avoid bias.
However, practical QPE has a finite energy resolution that is inversely proportional to the amount of Hamiltonian simulation time per QPE operation,
 causing a commensurate bias \cite{quantum_Metropolis_analysis}.
Furthermore, measurements of noncommuting observables collapse $\rho_\beta$
 and require both algorithms to re-equilibrate the quantum Markov chain for every sample of their thermal expectation values \cite{quantum_Metropolis_example}.

In this Letter, we recover passive, unbiased sampling of thermal expectation values by replacing the two principal components of a quantum Metropolis algorithm.
First, we sample the expectation value of an observable,
\begin{equation}
 L = \sum_{a} \lambda_a | \phi_a \rangle \langle \phi_a |,
\label{observable}
\end{equation}
 by directly measuring and repreparing its eigenstates $|\phi_a\rangle$
 to drive transitions between $|\psi_a\rangle$, which replaces unitary transition operations that extract no information from $\rho_\beta$.
Observable measurements thus occur passively inside the quantum Markov chain instead of as an external operation that breaks the chain.
Second, we incorporate a Gaussian filter into QPE operations (GQPE) that ideally measure an energy $\omega$ of an input state $| \Psi \rangle$
 while collapsing it into
\begin{equation}
 | \Psi(\omega) \rangle \propto \exp[ - (\omega - H)^2/(4\gamma)] | \Psi \rangle
\label{ideal_GQPE}
\end{equation}
 for some energy measurement variance $\gamma$.
This Gaussian form enables an exact bias correction,
 and we suppress all secondary sources of bias exponentially in the amount of Hamiltonian simulation time per GQPE operation.
While we cannot eliminate bias, we can efficiently reduce it well below the finite-sampling errors in expectation values.

For a quantum algorithm to establish a clear advantage over the best available classical algorithms for calculating
 thermal expectation values, it should be able to reduce its errors and increase its simulation size systematically and efficiently until reasonable doubts are satisfied.
A variety of algorithmic concepts have been proposed --
 variational free-energy minimization \cite{variational_free_energy},
 imaginary-time evolution \cite{QMETTS},
 coupling from the past \cite{coupling_from_the_past},
 artificial thermal baths \cite{bath_engineering},
 thermofield-double states \cite{thermofield_double},
 and local dissipative channels \cite{local_state_prep} --
 but only a few imaginary-time evolution algorithms \cite{sqrt_Gibbs_sampling,quantum_SVD,quantum_SDP_solver,quantum_function_evaluation}
 and the algorithm proposed here have established systematic error reduction for polylogarithmic costs.
An unraveling of $\rho_\beta$ as imaginary-time evolution of a maximally-mixed or maximally-entangled state for time $\beta/2$ is comparable to the
 $O(\beta)$ real-time evolution cost of our MCMC algorithm.
Since imaginary-time evolution is not unitary, it incurs a cost prefactor of $O(\sqrt{N})$ at large $\beta$
 for amplitude amplification, where $N$ is the Hilbert space dimension.
Our MCMC algorithm replaces this prefactor with mixing and stopping times that have controllable $N$ dependence.
Through brute-force classical simulation, we show that our MCMC algorithm can surpass the sampling
 efficiency of two representative non-MCMC algorithms as $N$ increases and biasing errors are kept under control.
To describe algorithms clearly and concisely, we write the high-level algorithms in pseudocode notation and the low-level quantum operations in standard quantum circuit notation.
The basic pseudocode operations are $x \leftarrow y$ to assign the value of $y$ to $x$
 and $z \sim P(z)$ to sample $z$ from the probability distribution or density $P(z)$,
 and classical variables are passed by value into functions.
We use kets to label quantum variables, which are passed by reference into functions since we cannot copy an arbitrary unknown quantum state.
The pseudocode contains two nonstandard function symbols -- $\theta(x)$ is the uniform distribution over $[0,1]$,
 and $\mathrm{POVM}(K_a, |\Psi\rangle)$ is a positive operator-valued measure
 that samples $a \sim \langle \Psi | K_a^\dag K_a |\Psi \rangle$ with the quantum side effect of $|\Psi\rangle \gets K_a |\Psi\rangle \langle \Psi | K_a^\dag K_a |\Psi \rangle^{-1/2}$.

\begin{figure}[!b]
\includegraphics{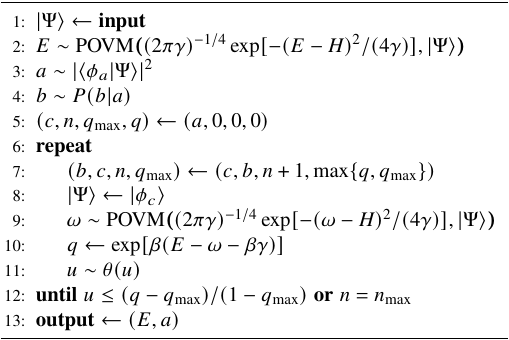}
\caption{Quantum Metropolis algorithm defined by a symmetric conditional probability distribution $P( b | a )$,
 which is unbiased in the $n_{\max} \rightarrow \infty$ limit for any energy measurement variance $\gamma$.
\vspace{-1.6mm}
}\label{pseudocode}
\end{figure}

Our quantum extension of the Metropolis algorithm in Fig.\@ \ref{pseudocode} mixes the input thermal state
 and outputs unbiased estimators $E$ and $a \mapsto \lambda_a$ of $\mathrm{tr}(\rho_{\beta} H)$ and $\mathrm{tr}(\rho_{\beta} L)$.
With $|\phi_a\rangle = |\psi_a\rangle$, it is equivalent to the classical algorithm in the $\gamma \rightarrow 0$ limit
 by stopping at $n=1$ upon accepting the trial state and $n=2$ upon rejection.
A common feature of our algorithm and the first quantum Metropolis algorithm \cite{quantum_Metropolis} is a repeat-until-success loop,
 but their loops differ in two important ways.
The previous loop was required only for rejection,
 but succeeded only by preparing a state with an energy equal to the input state.
Our loop has a broader success criteria that includes states with energies less than the input state,
 but it must alternate between the input and trial states to satisfy detailed balance for $\gamma > 0$.

\begin{figure}[!t]
\includegraphics{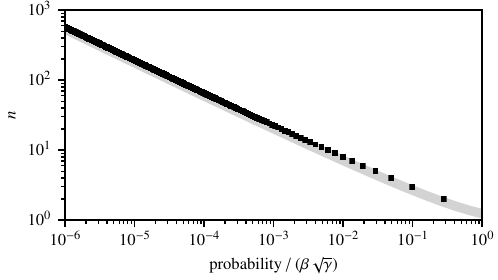}
\caption{Stopping-time distribution of Fig.\@ \ref{pseudocode} for a single-energy Hamiltonian at $\beta^2 \gamma \ll 1$,
 with limiting values of $1 - \beta \sqrt{\gamma/\pi}$ for $n = 1$ and $0.71 \beta \sqrt{\gamma / \log(n)} n^{-2}$ for $n \gg 1$ (gray line).
\vspace{-1.6mm}
}\label{stopping_distribution}
\end{figure}

Although it introduces bias, we impose a limit $n_{\max}$ on the number of iterations in each repeat-until-success loop
 to control stopping-time divergences caused by imprecise energy measurements.
Rare but large underestimates of $E$ will typically require an equally rare, large underestimate of $\omega$ to stop the loop.
The rarity of these events is offset by the amount of time required to resolve them.
A simple representative example of a bad stopping-time distribution is shown in Fig.\@ \ref{stopping_distribution} \cite{supplement},
 with a mean stopping time of
\begin{equation}
\mathbb{E}[n] \approx 1 + 1.4 \, \beta \sqrt{\gamma \log(n_{\max})} .
\label{stopping_mean}
\end{equation}
In general, this fat-tailed term is a fractional contribution to the average stopping time along with a system-specific term
 that depends on $H$, $L$, and $P(b|a)$, but not $n_{\max}$.

We analyze Fig.\@ \ref{pseudocode} as a map, $\rho \mapsto M(\rho)$, marginalized over $(E,a)$, unraveled over $n$, and in the $n_{\max} \rightarrow \infty$ limit,
\begin{align}
 & M(\rho) = \sum_{n=1}^{\infty} \sum_{a,b,c,d} M_{cb,n}^{da} | \psi_c \rangle \langle \psi_a | \rho | \psi_b \rangle \langle \psi_d | .
\label{map}
\end{align}
We further unravel this map over GQPE outcomes $\Omega_i$ on lines 2 and 9 of Fig.\@ \ref{pseudocode} by expanding its matrix elements as
\begin{align}
 M_{cb,n}^{da} = \int_{\mathbb{R}^{n+1}} \frac{d\Omega_n \cdots d\Omega_0}{(2 \pi \gamma)^{(n+1)/2}} A(\Omega_0^n) R(\Omega_0^{n-1}) S_{cb}^{da}(\Omega_0^n)
\label{energy_map}
\end{align}
 with notation $\Omega_n^m \coloneqq \Omega_m, \cdots, \Omega_n$ for multiple arguments.
Here, $A(\Omega_0^n)$ is the probability to accept iteration $n$, and
\begin{equation}
 R(\Omega_0^n) = \prod_{m \in \{0, \cdots, n \} \setminus \{ 0 \} } [1 - A(\Omega_0^m)]
\label{rejection_probability}
\end{equation}
 is the probability of rejecting all iterations up to $n$, where the empty set of products is evaluated as 1 by convention.
The remaining energy-filtered transition matrix elements,
\begin{align}
 &S_{cb}^{da}(\Omega_0^n) = e^{-(E_c - E_d)^2/(8\gamma)} e^{-(E_a - E_b)^2/(8\gamma)} \notag \\
 & \ \ \ \ \ \times e^{-[\Omega_n - (E_c + E_d)/2]^2/(2\gamma)} e^{-[\Omega_0 - (E_a + E_b)/2]^2/(2\gamma)}\notag \\
 & \ \ \ \ \ \times \sum_{g, f} P(g|f) \langle \psi_c | \phi_{h_n} \rangle \langle \phi_{h_n} | \psi_d \rangle \langle \psi_b | \phi_{h_0} \rangle \langle \phi_{h_0} | \psi_a \rangle \notag \\
 & \ \ \ \ \ \times \prod_{i \in \{1, \cdots , n \} \setminus \{ n \} } \sum_j e^{-(\Omega_i - E_j)^2/(2\gamma)} | \langle \psi_j | \phi_{h_i} \rangle|^2
\label{matrix_elements}
\end{align}
 for $h_{2i} \coloneqq f$ and $h_{2i+1} \coloneqq g$, are symmetric with respect to reversing the order of outcomes, 
 $S_{cb}^{da}(\Omega_0^n) = S_{bc}^{ad}(\Omega_n^0)$.

We now demonstrate that every loop iteration of Fig.\@ \ref{pseudocode} independently satisfies quantum detailed balance \cite{quantum_Metropolis},
\begin{equation}
  M_{c b, n}^{d a} e^{-\beta(E_a + E_b)/2} = M_{b c,n }^{a d} e^{-\beta(E_c + E_d)/2},
\label{detailed_balance}
\end{equation}
 to guarantee that $\rho_{\beta}$ is a stationary state.
First, we replace Boltzmann factors $e^{-\beta (E_a + E_b)/2}$ with measurable versions $e^{-\beta \Omega_i}$
 using a Gaussian summation identity,
 \begin{align}
 & \sum_{i \in \mathbb{Z}}  e^{- \beta E} f(\omega_i) e^{-(\omega_i - E)^2/(2 \gamma)} = e^{-\beta^2 \gamma/2}  \notag \\
 \times & \sum_{i \in \mathbb{Z}} e^{ - \beta \omega_i} f(\omega_i + \beta \gamma) e^{- (\omega_i - E)^2/(2 \gamma)}
\label{Gaussian_sum_identity}
\end{align}
 for $\omega_i = \omega_0 + i \beta \gamma / n$ and $n \in \mathbb{N}$,
 which applies to both the $\Omega_i$ integrals in Eq.\@ (\ref{energy_map}) and approximate GQPE operations with outcomes limited to a uniform energy grid.
Second, we enforce more fine-grained detailed balance on pairs of integrand values containing $S_{cb}^{da}(\Omega_0^n)$ and $S_{bc}^{ad}(\Omega_n^0)$,
 which enables a cancellation of matrix elements and reduction to
\begin{align}
  &A(\Omega_0^n + \beta \gamma) R(\Omega_0^{n-1} + \beta \gamma) e^{-\beta \Omega_0} = \notag \\
  &A(\Omega_n^0 + \beta \gamma) R(\Omega_n^1 + \beta \gamma) e^{-\beta \Omega_n}.
\label{detailed_balance2}
\end{align}
Thus, quantum detailed balance reduces to an observable constraint on the acceptance probability function $A$.

We satisfy Eq.\@ (\ref{detailed_balance2}) with a simple generalization of the standard Metropolis acceptance function to
\begin{equation}
  A(\Omega_0^n) = \min \left\{ \frac{R(\Omega_n^1 + \beta \gamma) e^{-\beta \Omega_n} }{R(\Omega_0^{n-1}) e^{-\beta (\Omega_0 - \beta \gamma)}}, 1 \right\}
\end{equation}
 that corresponds to a greedy maximization of each pair of acceptance probabilities.
By rewriting Eq.\@ (\ref{rejection_probability}) recursively to relate $R(\Omega_0^n)$ and $R(\Omega_0^{n-1})$,
 we can insert this choice of $A$ to evaluate $R(\Omega_0^n)$ for $n \ge 1$ explicitly as
\begin{equation}
 R(\Omega_0^n) = \max \left\{1 - \max_{m \in \{1, \cdots , n\}} e^{\beta ( \Omega_0 - \Omega_m - \beta \gamma)} , 0 \right\},
\end{equation}
 which enables the simple implementation of $A$ on lines 7, 10, and 12 of Fig.\@ \ref{pseudocode}.
The same acceptance function with $\gamma = 0$ has been used previously \cite{MCMC_rejection_delay} to delay the rejection process in the classical Metropolis algorithm.

\begin{figure}[!t]
\includegraphics{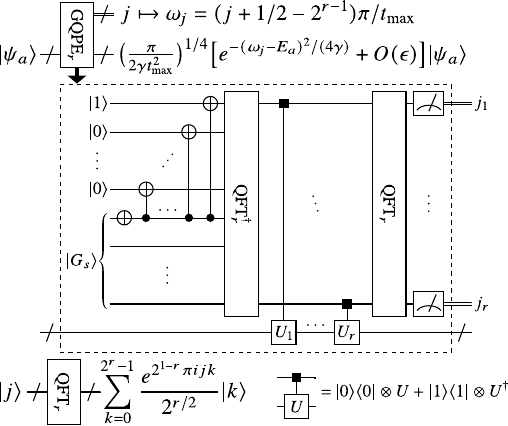}
\caption{Approximate GQPE operation using $r$ ancillary qubits, the $s$-qubit resource state $|G_s\rangle$,
 and Hamiltonian time evolution based on $U_j = e^{-i H t_{\max} 2^{-j}}$ with a maximum duration of $t_{\max}$.
\vspace{-1.6mm}
}\label{GQPE_circuit}
\end{figure}
 
To implement the GQPE operation, we incorporate the concept of quantum spectral filtering \cite{quantum_filter} into traditional QPE.
We begin by expanding Eq.\@ (\ref{ideal_GQPE}) into a measurement of an ancillary continuous quantum variable $\omega$,
\begin{align}
 |\Psi(\omega)\rangle \otimes |\omega \rangle &\propto \int_{-\infty}^{\infty} dt \frac{e^{i \omega t}}{\sqrt{2\pi}} g(t) e^{-i H t} |\Psi\rangle \otimes | \omega \rangle, \notag \\
 g(t) = \frac{e^{-\gamma t^2}}{[\pi/(2\gamma)]^{1/4}} &= \int_{-\infty}^{\infty} d\omega \frac{e^{-i\omega t}}{\sqrt{2\pi}} \frac{e^{-\omega^2/(4\gamma)}}{(2\pi \gamma)^{1/4}},
\label{expanded_GQPE}
\end{align}
that is prepared with a narrow Gaussian filter over $\omega$, then
 Fourier transformed into a wide Gaussian filter over $t$ and used as a control of Hamiltonian time evolution duration
 before its inverse Fourier transform back to $\omega$.
With only $r$ ancillary qubits available, we approximate the integrals over $\omega$ and $t$ in Eq.\@ (\ref{expanded_GQPE}) by finite sums over
\begin{equation}
 \omega_i = \omega_{\max} \frac{2 i + 1 - 2^r}{2^r}, \ \ \
 t_i = t_{\max} \frac{2 i + 1 - 2^r}{2^r},
\end{equation}
 for $i \in \{ 0, \cdots , 2^r-1 \}$.
If these sums are constrained by $\omega_{\max} t_{\max} = 2^{r-1} \pi$, then they can be implemented with the conventional quantum Fourier transform (QFT)
 as shown in Fig.\@ \ref{GQPE_circuit}.
We limit the total Hamiltonian simulation time to $t_{\max}$ using qubit-controlled Hermitian conjugation \cite{QPE_conjugate}
 and restrict the narrow Gaussian filter to an $s$-qubit state,
\begin{equation}
 | G_s\rangle \propto \sum_{j=0}^{2^s-1} e^{- i (1-2^{-r}) \overline{\omega}_j t_{\max} - \overline{\omega}_j^2/(4\gamma)} |j\rangle
\label{GQPE_resource}
\end{equation}
for $\overline{\omega}_i = \omega_{i + 2^{r-1} - 2^{s-1}}$.
The preparation of $| G_s\rangle$ is likely to require $O(2^s)$ gates \cite{quantum_prep},
 thus we must carefully choose $s$ alongside $r$ and $t_{\max}$ to balance cost and accuracy.

For the rest of this Letter, we restrict $\gamma$ to the specific choice of $\gamma = \pi / (\beta t_{\max})$
 that corresponds to the smallest energy variance compatible with Eq.\@ (\ref{Gaussian_sum_identity}).
This selection minimizes the average stopping time in Eq.\@ (\ref{stopping_mean}) associated with the generic stalling of the repeat-until-success loop in Fig.\@ \ref{pseudocode}.
It also causes the $\beta \gamma$ bias correction on line 10 of Fig.\@ \ref{pseudocode} to be a shift of just one energy grid point,
 which is the minimum separation between the input energy and an output energy that is guaranteed to be accepted.
Although this predetermined choice of $\gamma$ might not strictly minimize the overall resource usage,
 it has a substantial simplifying effect on the remaining cost and error analysis that aids in optimizing the other free parameters in the algorithm.

The main source of biasing error in our implementation of Fig.\@ \ref{pseudocode} is the approximation error on lines 2 and 9 from the output of Fig.\@ \ref{GQPE_circuit},
 which is a truncated Fourier series
\begin{equation}
 \tilde{G}(\omega) = \sum_{j=0}^{2^r - 1} \sum_{k=0}^{2^s - 1} \frac{e^{i (\omega - \overline{\omega}_k) t_j - \overline{\omega}_k^2/(4\gamma)}}{2^r} \approx e^{-\omega^2/(4\gamma)}
\end{equation}
 over the domain $|\omega| \le \Omega_{\max} = (1-2^{-r})\omega_{\max} + E_{\max}$ with $E_{\max} = \max_a |E_a|$.
We identify the leading-order Fourier errors and combine them into an empirical error bound,
\begin{align}
 &\max_{\omega \in [-\Omega_{\max}, \Omega_{\max}]} | e^{-\omega^2/(4\gamma)} - \tilde{G}(\omega) | \le \epsilon, \notag \\
 &\epsilon = 2 \exp [ - \min \{ \pi t_{\max}/\beta , \, 2^{2(s - 1)}\pi \beta / (4 t_{\max}) , \notag \\
 & \ \ \ \ \ \ \ \ \ \ \ \ \ (2^{r - 1} - E_{\max} t_{\max}/\pi )^2 \pi \beta / (4 t_{\max}) \} ],
\label{error_bound}
\end{align}
 contingent on $E_{\max} \le \omega_{\max}$, which we numerically verify over $r$, $s$, $t_{\max}$, and $E_{\max}$ \cite{supplement}.
This error bound allows us to minimize $r$, $s$, and $t_{\max}$ independently for a given $\epsilon$,
\begin{align}
 t_{\max} &= (\beta/\pi) \log(2/\epsilon),  \ \ \ s = \lceil \log_2 [ (4/\pi) \log(2/\epsilon)]\rceil, \notag \\
 r &= \lceil \log_2[  (2 \beta E_{\max} / \pi^2 + 4/\pi) \log(2/\epsilon) ] \rceil,
\label{resource_summary}
\end{align}
 relating GQPE resource estimates to $\beta$, $E_{\max}$, and $\epsilon$.

The accumulation of GQPE errors motivates the choice of $n_{\max}$.
We use the triangle inequality to apply Eq.\@ (\ref{error_bound}) to the product of approximate Gaussian filters in $M_{ca,n}^{db}$,
\begin{equation}
 \left| \exp\left(-\sum_{i=1}^{2n+2} \frac{\xi_i^2}{4\gamma}\right) - \prod_{i=1}^{2n+2} \tilde{G}(\xi_i) \right| \le (1+\epsilon)^{2n+2}-1 
\end{equation}
 for dummy variables $|\xi_i| \le \Omega_{\max}$.
These errors in $M_{ca,n}^{db}$ are avoided for $n > n_{\max}$, and they are instead replaced by the approximation errors from $A(\Omega_0^{n_{\max}}) \approx 1$.
We choose $n_{\max} = \lfloor 0.5 / \log_2(1+\epsilon) \rfloor - 1$, which is the crossover point between our simple bounds on these two sources of error.
The errors near the crossover point are large, but they are rare enough to have a small overall effect on the quantum map.
Errors in the map also induce errors in its stationary thermal state, which are further amplified by the stopping and mixing times of Fig.\@ \ref{pseudocode}.
Similar error propagation has been analyzed in more detail by others \cite{quantum_Metropolis_analysis},
 but we instead rely on empirical convergence of observables with respect to $\epsilon$ rather than detailed analysis because the mixing time
 of the quantum Markov chain is not known \textit{a priori}.

Having specified all of the parameters for our quantum Metropolis algorithm,
 we highlight an important property that simplifies its implementation.
While the algorithm in Fig.\@ \ref{pseudocode} defines a quantum Markov chain,
 we could instead partition subsequent Metropolis steps between lines 3 and 4,
 which defines a classical Markov chain having the state variables $(E,a)$.
Similarly, a quantum state does not have to be carried between iterations of the repeat-until-success loop,
 it only needs to be maintained long enough for one GQPE operation during the loop iterations and one more GQPE operation upon success.
When combined with low accuracy expectations and modest $t_{\max}$ values, this feature improves the viability of a quantum Metropolis algorithm
 on a noisy intermediate-scale quantum (NISQ) computing device \cite{NISQ} with circuit depths limited by noise.

\begin{figure}[!b]
\includegraphics{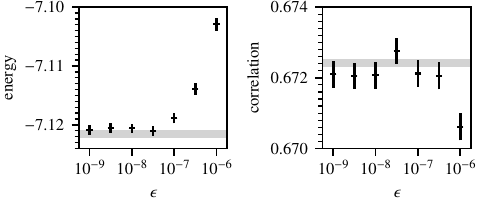}
\caption{Calibration of $\epsilon$ to reduce bias below standard errors for $H_8(\pi/4)$ in Eq.\@ (\ref{example_Hamiltonian}) and $\beta = 3$,
 driving expectation values from $10^6$ Metropolis samples towards their exact values (gray line).
\vspace{-1.6mm}
}\label{bias_calibration}
\end{figure}

We now apply our quantum Metropolis algorithm \cite{supplement} to a transverse-field Ising model on an $m$-site ring,
\begin{equation}
H_m(\theta) = - \sum_{i=1}^{m} [ \sin(\theta) X_i + \cos(\theta) Z_i Z_{1+i \, \mathrm{mod} \, m}],
\label{example_Hamiltonian}
\end{equation}
 where $X_i$ and $Z_i$ are Pauli operators on the $i$th qubit.
We perform measurements on the computational basis states, $|\phi_a\rangle = |a\rangle$ for $a \in \{0,1\}^m$,
 and apply a periodic sweep of spin flips defined by $P(x0y|x1y) = P(x1y|x0y) = 1$ with
 $x \in \{0,1\}^{i \, \mathrm{mod} \, m}$ and $y \in \{0,1\}^{m - 1 - (i \, \mathrm{mod} \, m)}$ corresponding to the $(i+1)$th Metropolis operation.
These observations allow us to sample local correlations, $\mathrm{tr}(\rho_\beta Z_i Z_{1+i \, \mathrm{mod} \, m})$.
As the magnitude of biasing errors is not known \textit{a priori}, we calibrate them by tuning $\epsilon$ as exemplified in Fig.\@ \ref{bias_calibration}.

To compare our quantum Metropolis algorithm against imaginary-time evolution, we consider two representative algorithms
 to serve as proxies for the previously proposed algorithms \cite{sqrt_Gibbs_sampling,quantum_SVD,quantum_SDP_solver,quantum_function_evaluation}.
The first algorithm performs a GQPE operation on a maximally-mixed state with a postselection probability $\min \{ 1, \exp[\beta (\omega_{\min} - \omega)] \}$ to sample from $\rho_{\beta}$ directly.
As is mostly the case in our quantum Metropolis algorithm, its circuits are only one GQPE operation deep,
 but postselection incurs an $O(N)$ cost prefactor at large $\beta$ in the Hilbert space dimension $N$.
The second algorithm applies amplitude amplification \cite{amplitude_amplification} to GQPE operations without ancillary measurements and maximally-entangled states,
 which balances the cost prefactor and circuit depth to $O(\sqrt{N})$.
In both algorithms, we must calibrate $\omega_{\min}$ to balance sampling efficiency with bias.
This pair of proxy algorithms enables cost comparisons in number of GQPE operations or duration of Hamiltonian time evolution per sample,
 which avoids further details of Hamiltonian oracle encoding and query complexity.
At low bias and large $\beta$, our two proxy algorithms have an $O(\beta)$ cost scaling in $\beta$,
 which is representative of previously proposed algorithms
 unless Hamiltonian fast-forwarding is utilized \cite{fast_forwarding,fast_forwarding_note}.

\begin{figure}[!t]
\includegraphics{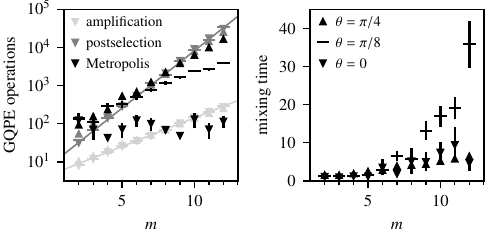}
\caption{Cost of a direct or Metropolis sample from the thermal state of $H_m(\theta)$ in Eq.\@ (\ref{example_Hamiltonian}) for $\beta = 3$ and $\epsilon = 10^{-8}$
 as quantified by number of GQPE operations.
Metropolis sample dependence is quantified by a mixing time that is a ratio of raw and effective sample sizes, estimated from $10^6$ Metropolis sample energies.
\vspace{-1.6mm}
}\label{sampling_rate}
\end{figure}

The efficiency of our quantum Metropolis algorithm is compared with the two proxy algorithms in Fig.\@ \ref{sampling_rate}.
Both proxy algorithms closely follow the guiding line for their asymptotic $O(\sqrt{2}^m)$ or $O(2^m)$ costs, independent of $\theta$.
In contrast, the Metropolis algorithm has an apparent $O(\alpha^{m})$ cost for $\alpha \in [1,2]$,
 which depends on the degree to which the Hamiltonian and measured observable commute.
The classical behavior of $\alpha = 1$ is recovered at $\theta = 0$ when $H$ and $L$ exactly commute,
 but $\alpha$ grows with increasing $\theta$ as the commutativity is reduced.
This cost-scaling behavior is similar to QPE-based ground-state preparation, which is
 independent of system size when the trial state has almost complete overlap with the ground state
 and exponential in system size when the overlap per subsystem is incomplete and approximately constant.
These state-preparation costs are related since an observable basis that commutes with a Hamiltonian 
 contains its ground state as an element.
 
In conclusion, we have constructed and demonstrated a practical quantum Metropolis algorithm
 with efficient bias reduction and passive observable sampling.
It possesses a low-depth circuit implementation and natural connections with QPE-based ground-state preparation.
While it retains an exponential system-size scaling of cost that is typical of quantum thermal-state preparation at low temperature and low bias,
 its growth of costs can be reduced by improving commutativity between observable and Hamiltonian.

\smallskip
J.\@ E.\@ M.\@ thanks Andrew Baczewski and Norm Tubman for helpful discussions.
The Molecular Sciences Software Institute is supported by NSF Grant No.\@ ACI-1547580.


\begin{thebibliography}{99}
\bibitem{fast_forwarding_note} One algorithm \cite{sqrt_Gibbs_sampling} has an improved $O(\sqrt{\beta})$ sampling cost, but it requires a square-root factorization of $H$ that enables a limited form of fast-forwarding \cite{fast_forwarding}.
\end{thebibliography}
\end{document}